\documentclass[aps,twocolumn,prc,floatfix,showpacs,preprintnumbers,amsmath,amssymb,
nofootinbib,groupedaddress]{revtex4-1}
\usepackage{wasysym}
\usepackage{cases}
\usepackage{mhchem}
\usepackage{multirow}
\usepackage{mathtools}
\usepackage{amsmath}
\usepackage{color}
\usepackage{graphicx}
\usepackage{dcolumn}    
\usepackage{multirow}
\usepackage{bm}         
\usepackage{sidecap}
\usepackage{ulem}
\usepackage{booktabs}			

\providecommand{\U}[1]{\protect\rule{.1in}{.1in}}

\newcommand{\be}{\begin{equation}}
\newcommand{\ee}{\end{equation}}
\begin{document}
%
\title{Outer crust of a cold non-accreting magnetar}

\author{D. Basilico\textsuperscript{1,2}, D. Pe\~na Arteaga\textsuperscript{3,4}, X. Roca-Maza\textsuperscript{1,2},} \email{xavier.roca.maza@mi.infn.it}\author{ G. Col\`o\textsuperscript{1,2}}

\affiliation{\textsuperscript{1} Dipartimento di Fisica, Universit\`a degli Studi di Milano, via Celoria 16, I-20133 Milano, Italy\\
             \textsuperscript{2} INFN, sezione di Milano, via Celoria 16, I-20133 Milano, Italy\\
             \textsuperscript{3} CEA, DAM, DIF, Bruy\`eres-le-Ch\^atel, F-91297 Arpajon Cedex, France\\
             \textsuperscript{4} Institute of Astronomy and Astrophysics, Universit\'e Libre de Bruxelles, CP 226, Boulevard du Triomphe, B-1050 Brussels, Belgium}

\date{\today} 

\begin{abstract}
   The outer crust structure and composition of a cold, non-accreting magnetar are studied. We model the outer crust to be made of fully equilibrated matter where ionized nuclei form a Coulomb crystal embedded in an electron gas. The main effects of the strong magnetic field are those of quantizing the electron motion in Landau levels and of modifying the nuclear single particle levels producing, on average, an increased binding of nucleons in nuclei present in the Coulomb lattice. The effect of an homogeneous and constant magnetic field on nuclear masses has been predicted by using a covariant density functional, in which induced currents and axial deformation due to the presence of a magnetic field that breaks time-reversal symmetry have been included self-consistently in the nucleon and meson equations of motion. Although not yet observed, for $B\gtrsim 10^{16}$G both effects contribute to produce different compositions --odd-mass nuclei are frequently predicted-- and to increase the neutron-drip pressure as compared to a typical neutron star. Specifically, in such a regime, the magnetic field effects on nuclei favor the appearance of heavier nuclei at low pressures. As $B$ increases, such heavier nuclei are also preferred up to larger pressures. For the most extreme magnetic field considered, $B=10^{18}$G, and for the studied models, the whole outer crust is almost made of ${}_{40}^{92}$Zr$_{52}$.  
\end{abstract}


\maketitle 


\section{Introduction}
\label{introduction}

Pulsars have typical surface magnetic fields of $10^{12}$ G, some displaying up to $10^{14}$ G. Larger magnetic fields on the surface of neutron stars have been observed or derived from observational data, with a maximum inferred value of $2.4 \times 10^{15} \, \mathrm{G}$ in soft gamma-ray repeaters (SGRs) and anomalous X-ray pulsars (AXPs) \cite{Seiradakis04,Ng10,Mereghetti08,Olausen14,Tiengo13}. Magnetic energy might be released in star-quakes producing short bursts of gamma-rays observed in SGRs and AXPs \cite{Duncan98}. The variation in the luminosity detected in such events suggests ${B\sim 10^{15} \, \mathrm{G}}$  \cite{Vietri07}. Stronger interior magnetic fields are thought to be present as suggested by various observations \cite{Stella05,Kaminker07,Rea10}. 

Based on theoretical calculations, the possible existence of magnetars displaying larger magnetic fields have been hypothesized by considering dynamo-like effects on the interior of very young neutron stars. Such effects might be enhanced by considering magnetic instabilities during the supernov\ae~ just before the neutron star is formed \cite{Thompson93,Woods,Ardeljan05}. According to the virial theorem and magneto-hydrodynamic simulations, the upper limit for the neutron-star interior magnetic fields is about $10^{18}$ G \cite{Lai91,Cardall00}. 

Matter properties in the outer envelopes of a magnetars are thought to be significantly modified by strong magnetic fields \cite{Haensel}. Magnetic fields alter transport processes and therefore transport properties, such as thermal and electrical conductivity \cite{Haensel,Potekhin96,Potekhin99,Ventura}. In addition, magnetic field stresses are thought to produce seismic modes that can be observed from Earth as quasi-periodic oscillations (QPOs) in the X-ray flux of giant flares from SGRs \cite{Carter06}.

For magnetic fields ${B\lesssim 10^{15} \, \mathrm{G}}$, the outermost layer of neutron star spans about seven orders of magnitude in density: from $10^4 \, \mathrm{g \, cm^{-3}}$, the complete ionization density, to approximately $4 \times 10^{11} \mathrm{g \, cm^{-3}}$ \cite{Haensel,Shapiro}, the neutron-drip density. In the simplest model, this outer crust is assumed to be made of nuclei and free electrons in their ground state at zero temperature \cite{Haensel}. Furthermore, we consider the system embedded in a uniform magnetic field. Within this typical density range, the ionized nuclei find energetically favorable to arrange themselves in a Coulomb lattice \cite{Shapiro}. In the lowest range of density, the energy associated to the electron gas and to the Coulomb crystal does not play a relevant role in the determination of the nuclear species present in the crust: thus, the most stable isotopes of nickel and iron will most likely show up in such conditions. As the density increases, the lattice energy effects remain negligible, while the energy of the electron gas raises significantly as compared to the total energy of the system. Such energy is lowered decreasing the electron number in the gas through electronic capture processes. These reactions imply a progressive neutron-enrichment for the nuclei in the lattice. Meanwhile this mechanism cannot go on indefinitely because of the growing nuclear symmetry energy. Thus the outer crust is the result of a competition between the electronic energy, favoring neutron-rich nuclei, and the nuclear symmetry energy, favoring fairly symmetric ones. The outer crust ends when the nuclei become unstable against neutron emission because of their high neutron imbalance, becoming the inner crust where a free neutron gas is also present \cite{chamel15a,chamel15b}. The ideal boundary between the two crusts is called neutron-drip transition, and the related dripping pressure $P_{\mathrm{drip}}$ plays a fundamental part in order to calculate the outer crust spatial extension \cite{Pearson12,Chamel12}.\\

Nuclei present in the Coulomb lattice might be very exotic. Rare Ion Beam Facilities worldwide aim at extending current mass measurements up to extreme values of neutron and proton asymmetries. However, accurate nuclear mass models predict the appearance of nuclei in the crust that have not been measured yet. The neutron star outer crust composition has been studied in various works, both in the absence \cite{Baym71,Ruester06,RocaMaza08} and presence of magnetic fields \cite{Lai91,Chamel12,pearson11,wolf13,kreim13}. Nevertheless, none of these previous analysis accounted for the effect of the external magnetic field on the nuclear binding energy of the nuclei present in the Coulomb lattice in a fully self-consistent manner. It will become evident in Sec.~\ref{results} how this dependence can change the ground state properties of matter in the crust. Specifically, we will rely on two commonly used covariant energy density functionals that have been successful in the global description of nuclear masses, charge radii, deformations and nuclear collective excitations. One of the models is based on the non-linear Walecka model \cite{Lalazissis97,boguta77,pannert87} and the other, more modern, on an effective Lagrangian with density dependent meson-nucleon couplings \cite{Lalazissis05}. 

The aim of this study is two-fold. First, to determine the composition of a cold non-accreting and strongly magnetized outer crust~(${B \geq 10^{14} \, \mathrm{G}}$) focusing on the differences on the composition and structure of the outer crust obtained when the effects of the magnetic field on nuclei are included or neglected. And second, to ascertain {\it if there are clear and model independent signatures} of these differences, and equally importantly, at what magnetic field magnitudes they become relevant. We will present our predictions up to the theoretically-based maximum magnetic field of $B\sim 10^{18} \, \mathrm{G}$, that guarantees mechanical stability to the star \cite{Lai91,Cardall00,Kiuchi08}.

The article is organized as follows. In Sec.~\ref{magnetar}, the formalism is briefly discussed. In Sec.~\ref{results}, we present the results for the structure and composition of the outer crust of a magnetar under the effect of strong magnetic fields. Our conclusions are laid in the last section.

\section{Outer crust in strong magnetic fields}
\label{magnetar}

We will highlight in this section the main features of our model for the neutron star outer crust, addressing the reader to \cite{Chamel12,PenaArteaga11} and references therein for further details.

The matter in the outer crust of a cold ($T=0$ K) non-accreting neutron star consists of a Coulomb lattice of completely ionized atoms (with proton number $Z$, neutron number $N$ and baryon number $A$) and a uniform Fermi gas of relativistic electrons \cite{Haensel, Shapiro}. Throughout the outer crust a continuous value of the pressure is required in order to ensure hydrostatic equilibrium; the matter density can suffer from discontinuities due to the change of the nuclei present in the Coulomb lattice. Hence, the Gibbs free energy per baryon $g(A,Z;P)$ at a constant pressure and zero temperature should be the thermodynamic potential to be optimized rather than the energy per particle $\varepsilon(A,Z;\rho)$ at constant matter density. At zero temperature 
\be g(A,Z;P)=\frac{E(A,Z;P)+PV}{A}=\varepsilon(A,Z;P)+\frac{P}{n} \label{gibbs0}\ee
where $V$ is the volume occupied by a unit cell of the Coulomb lattice, $n=A/V$ is the baryon density in such a cell and $\varepsilon(A,Z;P)=E(A,Z;P)/A$ is the corresponding energy per nucleon. The minimum of $g(A,Z;P)$ at a fixed pressure determines the couple $A$, $Z$ that provides the most stable configuration. The energy per baryon, $\varepsilon(A,Z;P)$, is composed of three terms, 
\be \varepsilon(A,Z;P)=\varepsilon_n(A,Z)+\varepsilon_e(A,Z;P)+\varepsilon_l(A,Z;P)\ , \label{ener}\ee
the {\it nuclear}, {\it electronic} and {\it lattice} energy terms per baryon, respectively. The nuclear term accounts in our model for the nuclear mass $M(A,Z)$, 
\be \varepsilon_n(A,Z)=\frac{M(A,Z)}{A}=\frac{Zm_p+(A-Z)m_n}{A}-\frac{BE(A,Z)}{A} \ee
where $m_p$ and $m_n$ are respectively the proton and neutron rest mass, $BE(A,Z)$ is the binding energy and we use, hereafter, natural units ($\hbar=c=1$). The nuclear energy does not depend on the density in a unit-cell since it is much smaller than the nuclear saturation density throughout the outer crust. Within our description, individual nuclei do not contribute to the total pressure which is composed of electronic and lattice contributions, respectively $P_e$ and $P_l$. It is customary to relate, via charge neutrality, the electronic Fermi momentum $p_{\mathrm{Fe}}$ with the baryon average density $n$ and define an average baryonic Fermi momentum $p_F$ as follows \cite{RocaMaza08},
\be p_{\mathrm{Fe}}= \left( 3\pi^2 n_e \right) ^{1/3}= \left( 3 \pi ^2 \frac{Z}{A} \, n \right) ^{1/3} \equiv \left( \frac{Z}{A} \right) ^{1/3} p_F \ee
where $n_e$ is the electron density. Such an explicit relation will be useful  analyzing the electronic and lattice energies. 

\subsection{Effects of the magnetic field on the electron gas}
\label{magnetar.electrons}

Extreme magnetic fields affect the electron gas energy and are able to change the outer crust structure and composition \cite{Chamel12}. The existence of the magnetic field requires a different treatment and, thus, a new analytical form of the $n_e$ and $g(A,Z;P)$ with respect to the $B=0$ case \cite{Chamel12,Lai91}. In the plane orthogonal to the magnetic field direction --we consider a uniform magnetic field $\vec{B}=B\hat{z}$, the electron motion is quantized into discrete Landau levels \cite{Rabi28}.  Assuming a relativistic electron Fermi gas embedded in an external uniform magnetic field, the Landau energy levels $E(\nu,p_z)$ can be written as follows \cite{Chamel12},
\be E^2(\nu,p_z)=p_z^2+m_e^2(1+2\nu B_{\star}) \quad \quad B_{\star}=\frac{B}{B_c} ~\label{landau} \ee
where $p_z$ is the electronic momentum along the magnetic field direction, $m_e$ is the electron rest mass, $E$ is the electron energy, $\nu$ is a non-negative  quantum number and $B_{\star}$ is the external magnetic field $B$ defined in units of the critical magnetic field $B_c$. The critical field is defined as the magnetic field at which the electron cyclotron energy equals the electron rest mass energy. That is,
$B_{c}=m_e^2/e \approx 4.41 \cdot 10^{13} \, \mathrm{G}$. 

Comparing Eq.~\eqref{landau} with the usual relativistic energy-momentum relation, a third additional energy term appears due to the electron interaction with the magnetic field.  This interaction energy is proportional to the quantum number $\nu$, and cannot exceed the electron chemical potential $\mu_e$. Hence the maximum number of Landau levels $\nu_{\mathrm{max}}$, related to the highest value of the interaction energy allowed between electrons and the external magnetic field, is evaluated setting $E(\nu_{\rm max},p_z=0)=\mu_e$ in Eq.~\eqref{landau}. This leads to the expression,
\be \nu_{\mathrm{max}}=\frac{1}{2 B_{\star}} \left( \frac{\mu_e^2}{m_e^2}-1 \right) \ee

From this equation, one sees that $\nu_{\mathrm{max}}$ is inversely proportional to the magnetic field. The latter is defined as \textit{strongly quantizing} if only the lowest level is filled. In the general case, the maximum electron momentum available, that is the Fermi momentum $p_{\mathrm{Fe}}$, can be computed setting $E(\nu,p_z\equiv p_{\mathrm{Fe}})=\mu_e$ in Eq.~\eqref{landau} for different $\nu$ values as
\be [p_{\mathrm{Fe}}(\nu)]^2+m_e^2 ( 1+2\nu B_{\star} )=\mu_e^2 \quad \quad 0 \leq \nu \leq \nu_{\mathrm{max}} \ee
It is customary to define the adimensional Fermi momentum $x_e(\nu)$ and the adimensional Fermi energy $\gamma_e$ as $ x_e(\nu)=p_{\mathrm{Fe}}(\nu)/m_e$ and $\gamma_e=\mu_e/m_e$, respectively. The electronic density $n_e$, energy $\varepsilon_e$, and pressure $P_e$ can be then calculated as \cite{Lai91}:
\be n_e=\frac{B_{\star}m_e^3}{4\pi^2}\sum_{\nu=0}^{\nu_{\mathrm{max}}} g_{\nu} x_e(\nu) \quad \quad g_{\nu}= \begin{dcases*}
        1 \quad \nu=0 \\
        2 \quad \nu \neq 0\\
        \end{dcases*} ~\label{ne} \ee
\be \varepsilon_e= \frac{B_{\star}m_e^4}{2\pi^2}\sum_{\nu=0}^{\nu_{\mathrm{max}}}{g_\nu(1+2\nu B_{\star}) \, \tau_+ \Big [ \frac{x_e(\nu)}{\sqrt{1+2\nu B_{\star}} } \Big ]} \ee
\begin{eqnarray} 
P_e&=& -n_e \varepsilon _e+n_e \mu_e \nonumber \\ 
   &=& \frac{B_{\star}m_e^4}{2\pi^2}\sum_{\nu=0}^{\nu_{\mathrm{max}}}{g_\nu(1+2\nu B_{\star}) \, \tau_- \Big [ \frac{x_e(\nu)}{\sqrt{1+2\nu B_{\star}} } \Big ] }
\end{eqnarray}
where,
\be \tau_{\pm}(x)=\frac{1}{2}x\sqrt{1+x^2} \pm \frac{1}{2}\ln{(x+\sqrt{1+x^2})} \ee

Unlike the $B=0$ case, these functions cannot be studied analytically, with the exception of the strongly quantizing magnetic field case. A detailed analysis can be found in \cite{Chamel12}. Finally, we note that we neglect the small electron exchange corrections \cite{pearson11}.

\subsection{Effects of the magnetic field on the binding energy of the nucleus}
\label{magnetar.nuclei}

Theoretical extrapolations on nuclear masses are required in order to describe the equation of state of the outer crust. Calculations of the needed nuclear binding energies employed in this work (both in absence and in presence of magnetic field) are based on the relativistic mean-field effective interactions NL3 \cite{Lalazissis97} and DD-ME2 \cite{Lalazissis05}. As mentioned, the former model corresponds to a non-linear Walecka model while the latter is based on an effective Lagrangian with density dependent meson-nucleon couplings. Differences between the models will allow us to assess --to some extent-- the model dependence in our results. Within the original and subsequent works, NL3 and DD-ME2 have been shown to be accurate in the description of experimental data on binding energies, charge radii, quadrupole deformations and the excitation energy of nuclear Giant Resonances \cite{Agbemava14}. 

The effects of the magnetic field on the binding energy of nuclei have been taken into account following the work of Ref.~\cite{PenaArteaga11}. For completeness, we highlight here the main features and address the reader to this work and references therein for further details. We have adopted a covariant density functional based on an effective Lagrangian with nucleons and mesons as the effective degrees of freedom~\cite{Gambhir90,Vretenar05}: 
\begin{equation}
\mathcal{L}=\mathcal{L}_{N}+\mathcal{L}_{m}+\mathcal{L}_{int}+\mathcal{L}%
_{BO}+\mathcal{L}_{BM};\label{e:lagdens}%
\end{equation}
where $\mathcal{L}_{N}$ refers to the Lagrangian of the free nucleon, $\mathcal{L}_{m}$ is the Lagrangian of the free meson fields and the electromagnetic field generated by the protons and $\mathcal{L}_{int}$ is the Lagrangian describing the interactions. These three terms compose the standard relativistic Lagrangian. Throughout this work, the parameter sets NL3~\cite{Lalazissis97} and DD-ME2~\cite{Lalazissis05} will be employed. These models differ on the form of $\mathcal{L}_{int}$. Specifically, for the case of NL3, 
\begin{align}
{\mathcal{L}}_{int}  &  =g_{\sigma}\bar{\psi}\sigma\psi-\frac{1}{3}g_2\sigma^3-\frac{1}{3}g_3\sigma^4\nonumber\\
   &-g_{\omega}\bar{\psi}\gamma_{\mu}\omega^{\mu}\psi\nonumber\\
   &  -g_{\rho}\bar{\psi}\gamma_{\mu}\vec{\tau}\vec{\rho}^{\mu}\psi\nonumber\\
   & -e\bar{\psi}\gamma_{\mu}A^{\mu}\psi\ , \nonumber
\end{align}
and for the case of DD-ME2,
\begin{align}
{\mathcal{L}}_{int} &  =g_{\sigma}\bar{\psi}\sigma\psi-g_{\omega}\bar{\psi}\gamma_{\mu}\omega^{\mu}\psi\nonumber\\
&  -g_{\rho}\bar{\psi}%
\gamma_{\mu}\vec{\tau}\vec{\rho}^{\mu}\psi-e\bar{\psi}\gamma_{\mu}A^{\mu}%
\psi\ ,\nonumber
\end{align}
where $e$ is the electric charge for protons, ${\psi}$ denotes the Dirac spinor, $\gamma_\mu$ the Dirac matrices, and $\vec{\tau}$ the Pauli matrices in isospin space. The meson-nucleon vertexes are denoted by ${g}_{i}$ for $i=\sigma$, $\omega$ and $\rho$; scalar-isoscalar, vector-isoscalar and vector-isovector fields, respectively. For the case of NL3, $g_i$ are constants and two non-linear terms have been introduced in the $\sigma$ field when compared to the case of DD-ME2. For the case of DD-ME2, the coupling constants are assumed to depend on the baryon density \cite{Lalazissis05}. The adopted ansatz for the density dependence has been guided by more fundamental Dirac-Brueckner-Harthree-Fock calculations in infinite nuclear matter. The total number of adjusted parameters to some selected experimental data are six for NL3 and eight for DD-ME2. 
 
In addition, there are  two terms corresponding to the interaction of the nuclear system with an external magnetic field. The coupling of the proton orbital motion with the external magnetic field, $\mathcal{L}_{BO}=-e\bar{\psi}\gamma^{\mu}A_{\mu}^{(\mathrm{ext})}\psi$ and the coupling of protons and neutrons intrinsic dipole magnetic moments with the external magnetic field \cite{Bjorken64} $\mathcal{L}_{BM}=-\bar{\psi}\chi_{\tau_{3}}^{(\mathrm{ext})}\psi,\label{BM}$ where $\chi_{\tau_{3}}^{(\mathrm{ext})}=\kappa_{\tau_{3}}\mu_{N}\frac{1}{2}\sigma_{\mu\nu}F^{(\mathrm{ext})\mu\nu}$, $F^{(\mathrm{ext})\mu\nu}$ is the external field strength tensor, $\sigma_{\mu\nu}=\frac{i}{2}\left[\gamma_{\mu},\gamma_{\nu}\right]$, $\mu_{N}=e\hslash/2m$ is the nuclear magneton and $\kappa_{n}=g_{n}/2$, $\kappa_{p}=g_{p}/2-1$ with $g_{n}=-3.8263$ and $g_{p}=5.5856$ being the intrinsic magnetic moments of protons and neutrons. Interactions with the external magnetic field are marked by the superscript $(\mathrm{ext})$. This field is considered to be externally generated, and therefore there is no associated field equation and thus no other bosonic terms in the Lagrangian. 

The magnetic field breaks spherical symmetry for the Dirac and Klein-Gordon equations \cite{PenaArteaga11}. Only axial symmetry is preserved. In addition, time-reversal symmetry is broken by the magnetic field leading to the appearance of time-odd mean fields and non-vanishing currents which induce space-like components of the vector mesons $\omega$ and $\rho$, usually referred as nuclear magnetism~\cite{Hofmann88,Koepf90,Afanasjev10}. 

The effects of the coupling of protons and neutrons to an external magnetic field can be classified as follows: \textit{nucleon  paramagnetism}, caused by the interaction of the magnetic field with the neutron (proton) magnetic dipole moment. Since the gyromagnetic factor for neutrons (protons) is negative (positive), configurations with the spin anti-parallel (parallel) to the magnetic field are energetically favored; and \textit{proton orbital magnetism}, caused by the coupling of the orbital motion of protons with the magnetic field. It favors configurations where the proton angular momentum projection is oriented along the direction of the external magnetic field. In general, it is thus expected that the magnetic field effects on the single-particle structure of nuclei are more pronounced for protons than for neutrons.

\begin{figure}
\includegraphics[width=1.0\linewidth,clip=true]{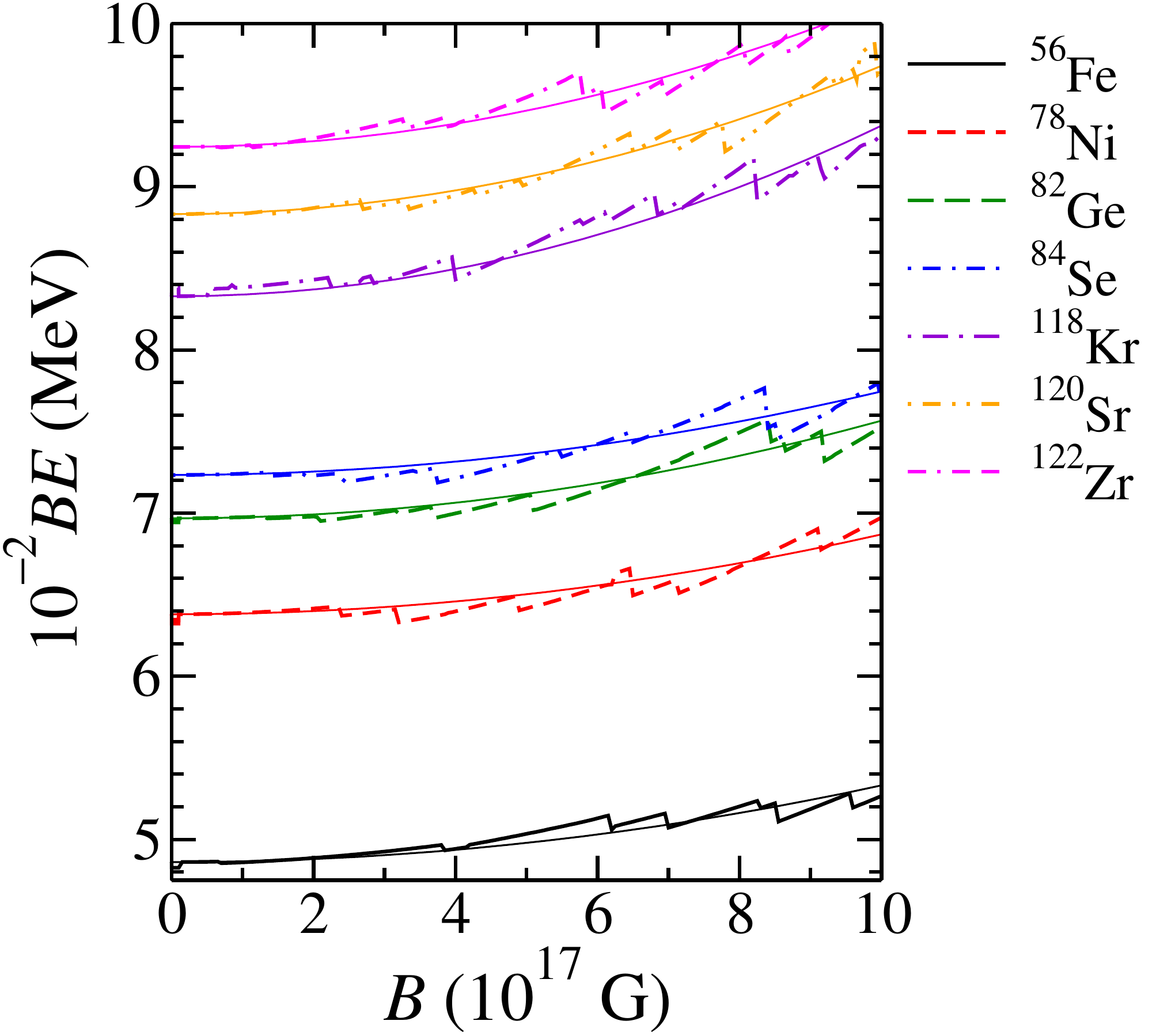}
\caption{(Color on-line) Binding energy $BE$ predicted by DD-ME2 as a function of the strength of the magnetic field $B$ for several typical nuclei appearing in the outer crust. A quadratic fit is also shown to indicate the average growing dependence of $BE$.}
 \label{fig:bind}
\end{figure}

From a more qualitative point of view, we show in Fig. \ref{fig:bind} the trends predicted by DD-ME2 in the binding energy $BE$, entering directly in the calculation of $M(A,Z)$, as a function of the external magnetic field for several typical nuclei thought to be present in the outer crust of a neutron star. From this figure, one sees on average a parabolic increasing trend of the binding energy with the magnetic field --for guidance, fitted parabolas are also shown\footnote{For a given unit cell ({\it i.e.} fixed volume) and a uniform $B$, the classical magnetic field energy scales with $B^2$}. We have checked that on average, the binding energy of nuclei present in the outer crust does not increase by more than a 10\% when the more extreme magnetic fields ($B\sim 10^{17-18}$ G) are taken into account. 

In the present work, we neglect the effect of pairing correlations on the nuclear binding energy, as it is expected that pairing effects will decrease with increasing magnetic field \cite{Kittel63}. Indeed, the interaction of the nucleus with the external magnetic field removes all degeneracies in the single-particle spectrum, and tends to separate formerly degenerate levels with opposing signs of angular momentum projection (cf. Figs. 2 and 3 of Ref.~\cite{PenaArteaga11}). Such  single-particle energy splitting will produce a reduction of the neutron and proton pairing gaps with increasing magnetic fields and, eventually, their disappearance \footnote{In analogy, the nuclear cranking model has been very useful for describing some transitional nuclei that behave as a {\it rotor}, thus, displaying a collective rotation of frequency $\omega_{\rm rot}$. In this context, it can be shown that for large values of the rotation frequency of about $\hbar \omega_{\rm crit}\approx 0.45$ MeV where the pairing gap disappears \cite{shimizu89}, the level degeneracy breaking also occurs. If one naively relates this rotation to be caused by an external magnetic field, one may find that $c B_{\rm crit}\approx \omega_{\rm crit} m c / e = \hbar\omega_{\rm crit} mc^2/\sqrt{\alpha}\approx 5\times 10^3$ MeV$^{2}$ which corresponds to about $10^{17}$ G.}.

\subsection{Effects of different lattice configurations}
\label{magnetar.lattice}

The lattice energy is not directly affected by the magnetic field \cite{VanVleck}. Nevertheless, some indirect effects on the lattice configuration due to Coulomb screening of the ions embedded in the electron gas may arise. The calculation of the potential energy of the Coulomb lattice\footnote{We would like to note that the quantum zero-point motion of ions~\cite{Baiko09} has been neglected throughout this work.} consists of divergent contributions that must cancel out as required by the overall charge neutrality of the system. Accurate calculations for the electron gas have been available for a long time and these results can be generalized to our case. Specifically, in the absence of a magnetic field, it has been shown that the most energetically favorable configuration is a crystallization into a body-centered cubic lattice \cite{Coldwell60, Fetter}. The lattice energy per baryon $\varepsilon_l(A,Z)$ may be written as \cite{Shapiro}, 
\be \varepsilon_l(A,Z,n)=-(1.81962)\frac{(Ze)^2}{a}=-C_{\mathrm{bcc}}\frac{Z^2}{A^{4/3}}p_F \ee
where $a$ is the lattice constant, $e$ is the electron charge, $p_F$ is in MeV and we have defined the coefficient ${C_{\mathrm{bcc}}=3.40665 \times 10^{-3}} $. The $\varepsilon_l$ dependence on the density $n$ or, equivalently, on the average baryonic Fermi momentum $p_F$, enters via the lattice constant $a$. Similar calculations can be carried out evaluating different lattice configurations, like faced-centered cubic or simple cubic ones: $\varepsilon_l$ dependencies on $A$, $Z$, $n$ hold and only $C_{\mathrm{bcc}}$ has to be replaced by $C_{\mathrm{fcc}}$ or $C_{\mathrm{sc}}$ \cite{Coldwell60}. The two latter coefficients are both smaller than $C_{\mathrm{bcc}}$, suggesting that a body-centered cubic lattice is the most favorable configuration in the absence of a magnetic field.\\

The body-centered cubic lattice pressure can be written as,
\be P_l(A,Z)=-\left. \frac{\partial E_l}{\partial V} \right| _{Z,A}=\frac{1}{3}\varepsilon_l=-\frac{n}{3}C_{\mathrm{bcc}}\frac{Z^2}{A^{4/3}}p_F \ . \label{pl}\ee

In the absence of a strong magnetic field, the effect on the lattice configuration due to the Coulomb screening of an ion embedded in a uniform electron gas can be neglected as far as the composition of the outer crust is concerned\footnote{Indeed, in the absence of a magnetic field and considering the specific conditions present in the outer crust, both the electron screening length $\lambda_e \approx \sqrt{\frac{\pi}{4\alpha}} p_{\mathrm{F}_e}^{-1}$ --within the mass-less Thomas-Fermi approximation \cite{Sharma10, Shalybkov87}-- and the ion-ion separation $r_\mathrm{ion-ion}=\frac{a}{\sqrt{2}}=\frac{1}{\sqrt{2}}\left(\frac{3Z}{2\pi}\right)^{1/3}p_{\mathrm{F}_e}^{-1}$ decrease with increasing electron densities. In particular, $\frac{r_\mathrm{ion-ion}}{\lambda_e} \approx \sqrt{\frac{2\alpha}{\pi}}\left(\frac{3Z}{2\pi}\right)^{1/3} \lesssim 0.2$ for the whole outer crust. This considerations justify the no-screening approximation.} \cite{Watanabe03, Pethick95}. In the presence of strongly quantizing magnetic fields such as those found in magnetars, a recent work \cite{Bedaque13} has shown that the most favorable lattice configuration for nuclei may not be the body-centered cubic lattice. Intense magnetic fields cause an anisotropic screening of the Coulomb force by the electron gas leading to Friedel oscillations in the ion-ion potential. Hence, for different values of the magnetic field, the authors of Ref.~\cite{Bedaque13} show that different Coulomb lattices such as faced-centered, hexagonal close-packed (hcp) or body-centered cubic oriented along the magnetic field could emerge. We have estimated that, for our purposes, the energy differences do not essentially affect the predicted structure and composition of the outer crust. On the other hand, it has been shown that, under certain conditions \cite{Jog82}, interpenetrating cubic lattices formed by different ions is energetically favourable with respect to a bcc lattice of any other single ion (assuming a uniform electron gas and in the absence of a magnetic field). Qualitatively, the same features are expected to remain when anisotropies of the background electron gas appear due to, for example, intense magnetic fields. Therefore, we have neglected the small energy correction due to such effects.

Among the three terms in Eq.~(\ref{ener}), the lattice configuration --also considering screening, anisotropies or other secondary effects-- do not significantly affect the crust composition. We verified that employing different lattice configurations or even neglecting the lattice for different magnetic values, the composition obtained is very similar with respect to the body-centered cubic case. In this regard the lattice configuration plays a very minor role and therefore from now we will simply treat nuclei as vertexes of a body-centered cubic lattice.

For a fixed uniform magnetic field and pressure, the calculation of the Gibbs energy per particle can be now explicitly written as, \cite{Shapiro}
\be g(A,Z;P,B)= \frac{M(A,Z;B)}{A}+\frac{Z}{A}\Big( \mu_e(P)+4\frac{P_l(A,Z)}{n_e} \Big) ~\label{gibbs} . \ee
and one can search for the optimal nucleus, solving the set of equations that determine $\mu_e$, $\nu_{\mathrm{max}}$, $n_e$, $p_{\mathrm{Fe}}(\nu)$:
\be
 \begin{dcases*}
        \mu_e^2=m_e^2\Big( 1+2\nu_{\mathrm{max}} B_{\star} \Big) \\
        p_{\mathrm{Fe}}(\nu)^2+m_e^2 \Big( 1+2\nu B_{\star} \Big) = \mu_e^2 \quad  0\leq \nu \leq \nu_{\mathrm{max}} \\
       n_e=\frac{B_{\star}m_e^3}{2 \pi^2} \sum_{\nu=0}^{\nu_{\mathrm{max}}}{g_\nu x_e(\nu)} \\
        P=P_e+P_l(A,Z) \\
        \end{dcases*} \label{system}
\ee

\subsection{The neutron-drip transition point} 
In this subsection we schematically study, using a toy model, how the pressure at the neutron-drip point changes in the presence of extreme magnetic fields ($B\gtrsim 10^{16}$G) which has both a direct effect on the pressure, and an indirect one through changes in the nuclear binding energies --nuclei do not contribute to the pressure in our model but determine the electron chemical potential at the bottom layer of the outer crust. For more details on the effects on the magnetic field on the neutron-drip transition point, we address the reader to Refs.\cite{chamel15b,chamel15a}. In these works, where the magnetic field effect on nuclei was neglected, it was found that $P_\mathrm{drip}$ increases linearly with extreme magnetic fields ($B\gtrsim 10^{16}$G).

To develope a simple, yet physical, model of the neutron-drip transition pressure $P_{\mathrm{drip}}$, we will neglect the small lattice contribution to the Gibbs free energy per baryon and, consistently also to the pressure. This assumption applied into the neutron-drip transition point allows us to write Eq.(\ref{gibbs}) as follows,
\be
m_n \approx \frac{Z_{\mathrm d}}{A_{\mathrm d}}m_p + \frac{N_{\mathrm d}}{A_{\mathrm d}}m_n - \frac{BE_{\mathrm d}}{A_{\mathrm d}} +\frac{Z_{\mathrm d}}{A_{\mathrm d}}\mu_{e\mathrm{, d}}  
\label{muedrip}
\ee
where the subscript 'd' denotes that the quantity should be evaluated at the neutron-drip transition. After some algebra, one finds, $\mu_{e\mathrm{, d}} \approx m_n - m_p + BE_{\mathrm d}/Z_{\mathrm d}$ where the neutron to proton mass difference can be neglected at this approximation level.

Based on the recent results found in Ref. \cite{chamel15b} and within our approximations, one may write the following expression for the dripping pressure,
\be
P_{\mathrm{drip}}\approx \frac{B_\star \mu_{e\mathrm{, d}}^2 m_e^2}{4\pi^2}\approx\frac{B_\star  m_e^2}{4\pi^2}\frac{BE_\mathrm{d}^2}{Z_\mathrm{d}^2} \ .
\label{pdrip}
\ee  
that is valid only for extreme magnetic fields ($B\gtrsim 10^{16}$ G). Eq.(\ref{pdrip}) suggest that $P_\mathrm{drip}$ changes linearly with the magnetic field and quadratically with the binding energy of the drip nucleus, which, at the same time, increases with increasing magnetic field (cf. Fig.\ref{fig:bind}). Specifically, we expect from this formula together with the average results shown in Fig.\ref{fig:bind} that the increase of $P_\mathrm{drip}$ between $B= 10^{16}$ G and $B= 10^{17}$ G should be linear with $B_\star$ and increase (roughly) one order of magnitude since the binding energy in nuclei is just barely affected on average by the strong magnetic field \cite{chamel15b}. The situation we expect between $B= 10^{17}$ G and $B= 10^{18}$ G is similar since the binding energy increases on average by a few \% (no more than a $\sim$ 10\%, cf. Fig.~\ref{fig:bind}). Therefore, $P_\mathrm{drip}$ should increase according to Eq.~(\ref{pdrip}) by one order of magnitude corrected by a small factor due to the increase in $BE$.

\begin{figure}[t!]
\includegraphics[width=1.0\linewidth,clip=true]{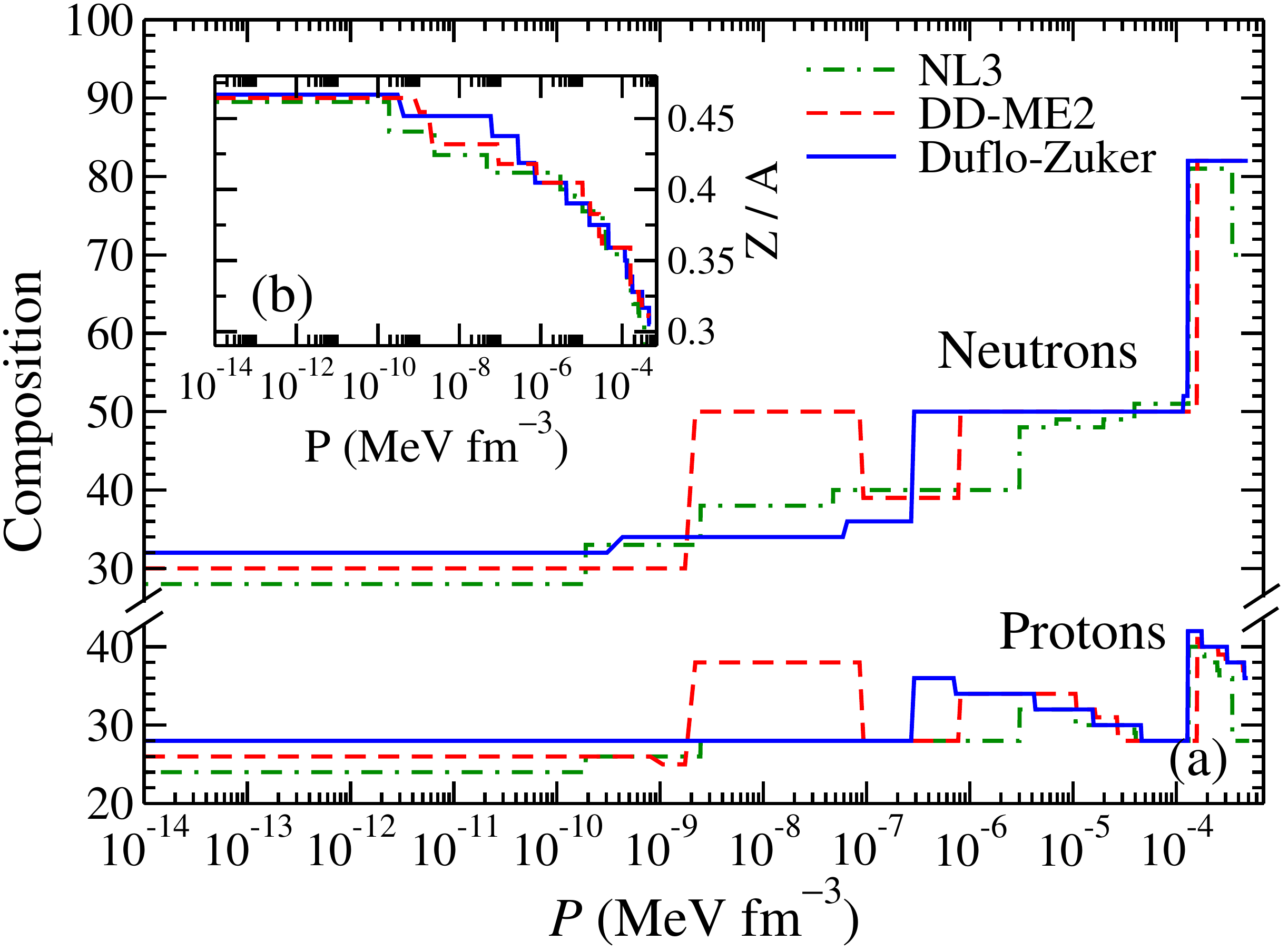}
\caption{(Color on-line) Proton and neutron trends (a) and proton fraction $y(P)\equiv Z(P)/[N(P)+Z(P)]$ trends (b) predicted by Duflo-Zuker model (blue solid line), by DD-ME2 model (red dashed line) and by NL3 model (green dash-dotted lines), in absence of magnetic field.}
\label{fig:zero}
\end{figure}

\section{Results}
\label{results}

We present in this section the main results obtained from the numerical minimization of the Gibbs energy per particle, $g(A,Z; P, B)$, from the outermost part of the outer crust (ionization at a pressure $P_{\rm ion}$\footnote{The ionization pressure $P_{\rm ion}$ corresponds, in good approximation, to the free Fermi electron density at which the electron binding to the nucleus $BE_{\rm elec}$ is not any more favorable. That is, $BE_{\rm elec}(Z)=\frac{3}{5}Z\varepsilon_F^{\rm ion}$. At low pressure (density) the most favorable nucleus has $Z\sim 26-28$, this implies a $P_{\rm ion}\sim 10^{-15}$ MeV fm$^{-3}$.}) to the innermost part (neutron-drip transition at a pressure $P_{\mathrm{drip}}$\footnote{The neutron-drip transition is determined by the condition $g(P_{\mathrm{drip}})=m_n$: it corresponds to the condition in which it is energetically favorable for the system to start dripping neutrons from the nucleus and form a neutron gas.}). Beyond the neutron-drip transition, where the inner crust begins, the presence of the neutron gas requires the addition of an extra term in the Gibbs energy \cite{Shapiro,Lai91,nandi11}. We focus our study on the outer crust composition as a function of the pressure $P(n)$, paying special attention to the effects produced by strong magnetic fields. To understand the importance of the change in binding energy of nuclei induced by the magnetic field, results including ($B_{\rm nucl}=B$) and excluding its effects (${B_{\rm nucl}=0 \, \mathrm{G}}$) on the composition will be presented alongside. The effect of the magnetic field on electrons is always taken into account.

\begin{figure}[t!]
\includegraphics[width=1.0\linewidth,clip=true]{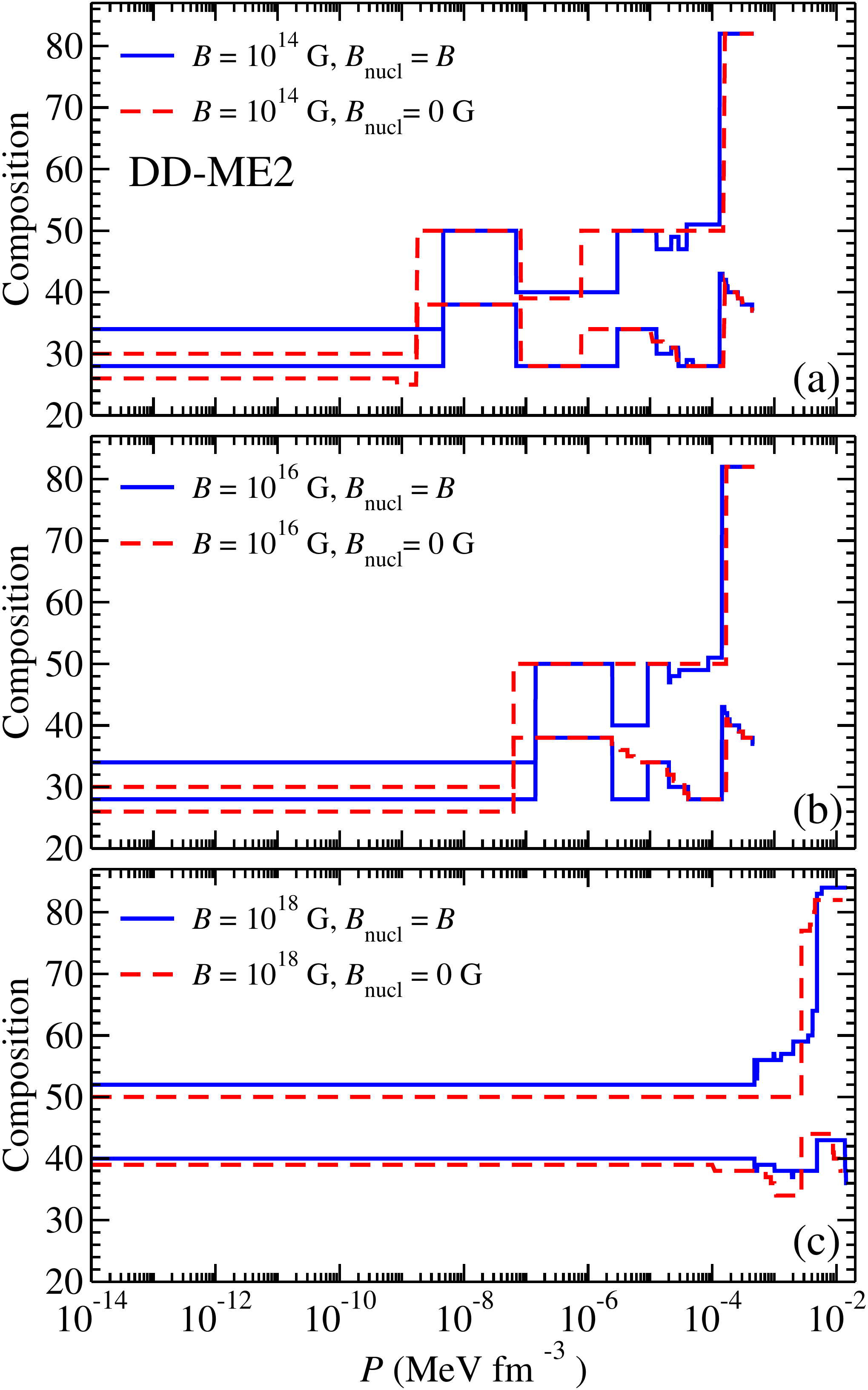}
\caption{(Color on-line) Compositions, i.e. $Z(P)$ and $N(P)$ trends, of the outer crust of a magnetar employing DD-ME2 model for three external magnetic fields $B$ values: (a) ${B=10^{14} \, \mathrm{G}}$, (b) $B=10^{16} \, \mathrm{G}$, (c) $B=10^{18} \, \mathrm{G}$. Solid (blue) lines include effects of magnetic field on nuclei and electrons, i.e. $B_{\rm nucl}=B$. Dashed (red) lines consider only magnetic effects on electron gas only, i.e. $B_{\rm nucl}=0 \, \mathrm{G}$. Note that upper lines correspond to $N(P)$ and lower lines to $Z(P)$.}
 \label{fig:ddme2}
\end{figure}

\begin{figure}[t!]
\includegraphics[width=1.0\linewidth,clip=true]{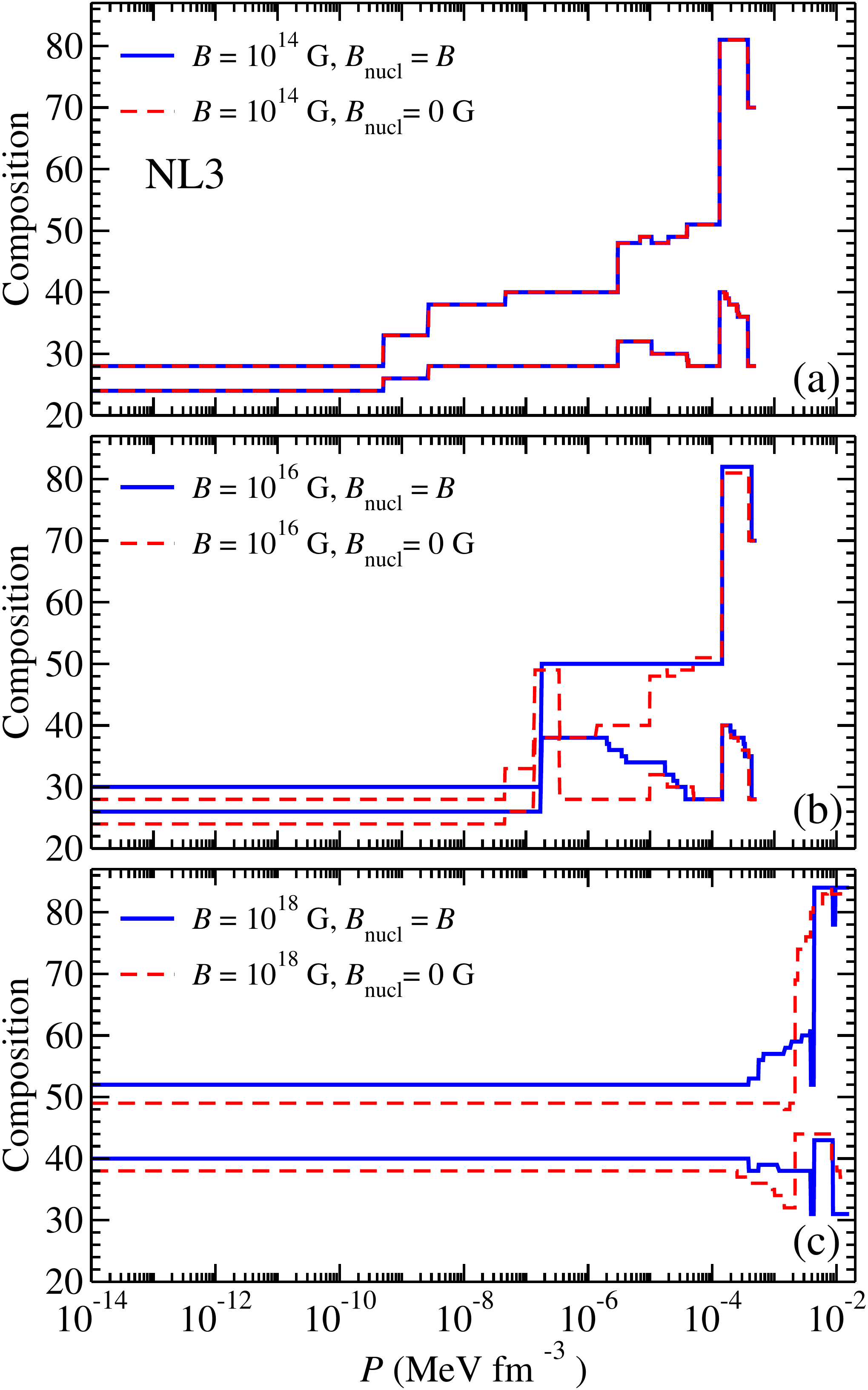}
\caption{(Color on-line) Compositions, i.e. $Z(P)$ and $N(P)$ trends, of the outer crust of a magnetar employing NL3 model for three external magnetic fields $B$ values: (a) $B=10^{14} \, \mathrm{G}$, (b) $B=10^{16} \, \mathrm{G}$, (c) $B=10^{18} \, \mathrm{G}$. Solid (blue) lines include effects of magnetic field on nuclei and electrons, i.e. $B_{\rm nucl}=B$.. Dashed (red) lines consider only magnetic effects on electron gas only, i.e. $B_{\rm nucl}=0 \, \mathrm{G}$. Note that upper lines correspond to $N(P)$ and lower lines to $Z(P)$.}
 \label{fig:nl3}
\end{figure}

If nuclear binding energies are assumed not to be affected by the external magnetic field (${B_{\rm nucl}=0 \, \mathrm{G}}$), one expects to find the same nuclear composition than in the ${B=0 \, \mathrm{G}}$  case with the only difference that the transitions from one nuclear species to another occur at higher pressures \cite{Chamel12,Lai91}. The latter effect is due to the electron interaction with the magnetic field: it reduces the Fermi energy $\mu_e$ and therefore delays, in terms of pressure, the appearance of a new nuclear species. It is important to note that for extremely large magnetic fields, of the order of $B\sim 10^{16-18}$ G, the effects of the electron gas can also change the composition \cite{Chamel12}. If one also accounts for the effects of a strong magnetic field on the nuclear binding energies (${B_{\rm nucl}=B\ne 0}$), further changes in the composition and discrepancies in the appearance of the subsequent nuclear species might be expected. As discussed in Sec.~\ref{magnetar.nuclei}, as $B$ increases, nuclei display larger binding energies on average (see Fig.~\ref{fig:bind}). This induces the appearance of different nuclei populating the Coulomb lattice. 

To check the qualitative behavior of the NL3 and DD-ME2 models, in Fig.~\ref{fig:zero}a we compare their optimal proton $Z(P)$ and neutron $N(P)$ numbers as a function of pressure in the absence of magnetic field with the predictions of the Duflo-Zuker model \cite{Duflo94,Zuker94,Duflo94b}, one of the most accurate mass models available in the literature (root mean square deviation of 400 keV). We notice that NL3 and DD-ME2 follow the main trends predicted by Duflo-Zuker\footnote{DD-ME2 displays a dip in the neutron and proton number between $P\sim10^{-7}-10^{-6}$ MeV fm$^{-3}$ and NL3 displays a gradual change in the neutron number as the pressure increases. Both are expected behaviors due to the fact that we neglect pairing correlations. As mentioned, we are interested on the effect of very high magnetic fields on the outer crust where pairing effects are expected to be reduced.}. That is, two main changes in the neutron number to a two stable plateaus and, in coincidence, a clear change on the proton number with a subsequent decrease due to electron capture processes. Looking in more detail, the main differences arise in the transition of the neutron and proton numbers at pressures between $10^{-9}$ and $10^{-6}$ MeV fm$^{-3}$. Depending on the symmetry energy predicted by each model at an average nuclear density, the transitions between the different neutron plateaus may appear at very different pressures \cite{RocaMaza08}: the larger the symmetry energy, the earlier the transition takes place. Hence, DD-ME2 shows the larger symmetry energy at an average nuclear density, then Duflo-Zuker and finally NL3. Although the models used are accurate in the description of stable nuclei, the model dependence just seen in the results at $B=0$ G will translate to the $B\ne 0$ G cases and will allow us to estimate, to a sizable extent, the model dependence of our results. In the inset of the same figure, Fig.~\ref{fig:zero}b, the proton fraction $y(P)\equiv Z(P)/[N(P)+Z(P)]$ is displayed, highlighting their good agreement. 

In Fig.~\ref{fig:ddme2} and Fig.~\ref{fig:nl3} are displayed the compositions,  i.e. $Z(P)$ and $N(P)$ trends, for three selected external magnetic field values $B$, employing respectively the DD-ME2 and NL3 models: (a) $B=10^{14} \, \mathrm{G}$, (b) $B=10^{16} \, \mathrm{G}$, (c) ${B=10^{18} \, \mathrm{G}}$. Solid (blue) lines include effects of magnetic field on nuclei and electrons, i.e. $B_{\rm nucl}=B$. Dashed (red) lines consider only magnetic effects on electron gas, i.e. $B_{\rm nucl}=0 \, \mathrm{G}$ ---note that upper lines correspond to $N(P)$ and lower lines to $Z(P)$. 

We focus first on the DD-ME2 results for $B_{\rm nucl} = 0$ G --dashed (red) lines in Fig.\ref{fig:ddme2}--, that is, only electrons are assumed to feel the magnetic field effects. We see in Fig.\ref{fig:ddme2}a that the changes in the neutron plateaus --from $N=28$ to $N=50$ and then to $N=82$-- are essentially unchanged with respect to the $B=0$ G case. The same happens for protons, they follow the same trend. Regarding the composition, it remains very similar to the $B=0$ G case depicted in Fig.~\ref{fig:zero}. In the same panel, we then see, that turning on the effects of the magnetic field on nuclei --solid (blue) lines, the appearance of the low pressure neutron plateau is shifted towards higher pressures, leaving unchanged the high pressure one. In addition, the composition is slightly changed at the lower pressures shown in Fig.\ref{fig:ddme2}~\footnote{These changes are thought to be model dependent since we will see that NL3 predicts no change.}.  

In Fig.\ref{fig:ddme2}b, there is a clear shift of the neutron plateau $N=50$ to higher pressures in both cases ($B_{\rm nucl}=0$ G and $B_{\rm nucl}=B$). The results on $B_{\rm nucl}=0$ G are therefore in agreement with previous literature \cite{Chamel12,Lai91}. The predicted composition is similar in the two cases --except for a dip in the case in which $B_{\rm nucl}$ is active-- and similar to the $B=0$ G case. Again, the main difference regarding the composition is on the low-pressure regime and due to the effect of the magnetic field on the binding energy of nuclei.

For the strongest magnetic field, $10^{18}$ G, shown in Fig.\ref{fig:ddme2}c, the situation is different. The structure between the $B_{\rm nucl}=0$ G and $B_{\rm nucl}=B$ cases is very similar and clearly different to the $B=0$ G. One may say that just one nucleus is composing the outer crust for almost the whole range of pressures. Regarding the composition, it is now clear that including the magnetic field effects on nuclei will be very important for a precise understanding of the outer crust in such conditions. Specifically we found ${}_{40}^{92}$Zr$_{52}$ to be the nucleus in such a constant plateau if the magnetic field effects on nuclei are taken into account.

In order to estimate the model dependence of the results we have just presented in Fig.\ref{fig:ddme2}, we show in Fig.\ref{fig:nl3} our results using a different model, the so called NL3. In Fig.\ref{fig:nl3}a, we show the results for $B=10^{14}$ G, there are no essential differences between the $B_{\rm nucl}=0$ G and $B_{\rm nucl}=B$ cases and with respect to the case in which the magnetic field effects on both electrons and nuclei are completely neglected (cf. Fig.~\ref{fig:zero} for $B=0$ G). 

Regarding the $B=10^{16}$ G predictions, shown in Fig.\ref{fig:nl3}b, we see that the composition in the low pressure regime is different when compared to Fig.\ref{fig:nl3}a, for both $B_{\rm nucl}=0$ G and $B_{\rm nucl}=B$. The $N=50$ plateau appears shifted to higher pressures also in both cases. The case in which $B_{\rm nucl}=0$ G --red (dashed) lines-- follows the same trends as the $B_{\rm nucl}=B$ G --blue (solid) lines-- except for a region in which a dip in the composition is found. This region spans almost two orders of magnitude in pressure, from $3\times 10^{-7}$ MeV fm$^{-3}$ to about $10^{-5}$ MeV fm$^{-3}$. This behavior has been also seen in Figs.~\ref{fig:zero} and \ref{fig:ddme2}b, the details of which are model dependent. 

\begin{figure}[Ht]
\includegraphics[width=1.0\linewidth,clip=true]{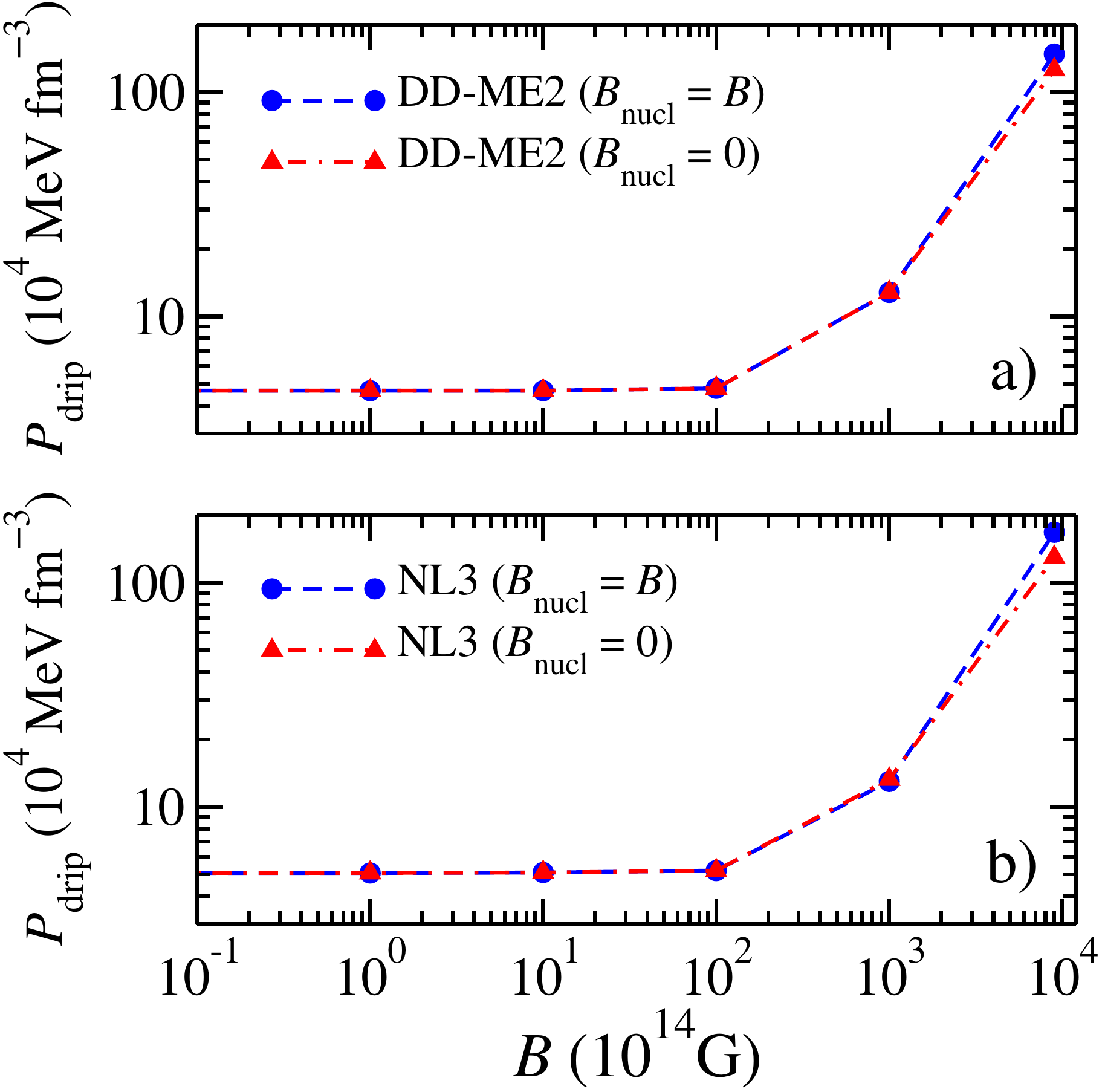}
\caption{(Color on-line) Neutron-drip transition pressures $P_{\mathrm{drip}}$ as a function of five external magnetic field \(B\): $B \approx 0 \, \mathrm{G}$, $B=10^{14} \, \mathrm{G}$, $B=10^{15} \, \mathrm{G}$, $B=10^{16} \, \mathrm{G}$, $B=10^{17} \, \mathrm{G}$ and  $B=10^{18} \, \mathrm{G}$. Circles (blue) refer to the results considering the effects of the magnetic field on both electrons and nuclei, i.e. \(B_{\rm nucl}=B\), while triangles (red) refers to calculations where the effects on nuclear binding energy have been neglected, i.e. \(B_{\rm nucl}=0\).}
 \label{fig:drip}
\end{figure}

For the highest magnetic field considered here, shown in Fig.\ref{fig:nl3}c, we found the same type of behavior as in Fig.\ref{fig:ddme2}c. That is, there is a constant plateau of neutron and proton numbers for almost the whole range of pressures relevant for the outer crust. Considering the effects of the external magnetic field on the binding energy of nuclei, turns out to be important. Specifically, we found ${}_{40}^{92}$Zr$_{52}$ again to be the nucleus in such a constant plateau for NL3 if the magnetic field effects on nuclei are taken into account. We have to emphasize that there is nothing special about this nucleus or in the fact that both models predict the same, being the change in single-particle level scheme the driving mechanism of the model dependence on the magnetic field, and considering that, for stable mid-mass nuclei, the level schemes are very similar for both DD-ME2 and NL3. In addition, it is expected that models with differing effective masses will yield a different nucleus sitting at the lattice.
      
Summarizing, for field strengths of about $10^{14}$ G the inclusion of the effect of the magnetic field on nuclear binding energies may play a role (depending on the nuclear mass model used, cf. Fig.~\ref{fig:ddme2}a and Fig.\ref{fig:nl3}a), and cannot be neglected for higher field strengths. If prominent magnetic fields, $B=10^{16-18} \, \mathrm{G}$, are confirmed to exist in the the surface of neutron stars, one should expect some changes in the structure and composition of the outer crust compared to the zero magnetic field case: 
\begin{itemize}
\item The effect of the magnetic field on electrons shifts to higher pressures the appearance of a new nuclear species, keeping almost unchanged the nuclei that populates the Coulomb lattice \cite{Chamel12,Lai91}.   
\item The magnetic field effect on nuclei favors heavier nuclei at lower pressures. As $B$ increases, such heavier nuclei are also preferred up to higher pressures. In the most extreme case, and for the studied models, the whole outer crust is almost composed of ${}_{40}^{92}$Zr$_{52}$, regardless of the nuclear model used to calculate the nuclear binding energies.
\item Extreme magnetic fields also favor in some cases the appearance of odd-mass nuclei in the outer crust, in contrast to what happens for lower magnetic fields where the pairing interaction has a more relevant role.
\end{itemize}

In Fig.~\ref{fig:drip} we display the evolution of the neutron-drip transition pressure $P_{\mathrm{drip}}$ as a function of the external magnetic field. Circles (blue) refer to the results including the effects of the magnetic field on both electrons and nuclei, i.e. \(B_{\rm nucl}=B\), while triangles (red) indicate that the effects on nuclear binding energy have been neglected, i.e. \(B_{\rm nucl}=0\). For the strongest fields, $10^{17-18}$G, the neutron-drip transition occurs, respectively, at a pressure one and two orders of magnitude higher than for weaker field strengths, where the $P_{\mathrm{drip}}$ coincides with the results for zero magnetic field. The effect of the magnetic field on nuclei produces just a small change in $P_{\mathrm{drip}}$. Our results are in agreement with Refs.\cite{Chamel12, chamel15b, chamel15a} since a change by an order of magnitude in $B$ for the most extreme cases, induces a change of one order of magnitude in $P_{\mathrm{drip}}$, that is, (roughly) a linear change. It is not surprising, with Fig.\ref{fig:bind} and Eq.(\ref{pdrip}) on hand, we could have expected the results shown in Fig.\ref{fig:drip}: changes on $P_{\mathrm{drip}}$ induced by the effect of the magnetic field on nuclear binding energies are overshadowed by changes induced by the magnetic field on the electrons.

\begin{figure}[Ht]
\includegraphics[width=1.0\linewidth,clip=true]{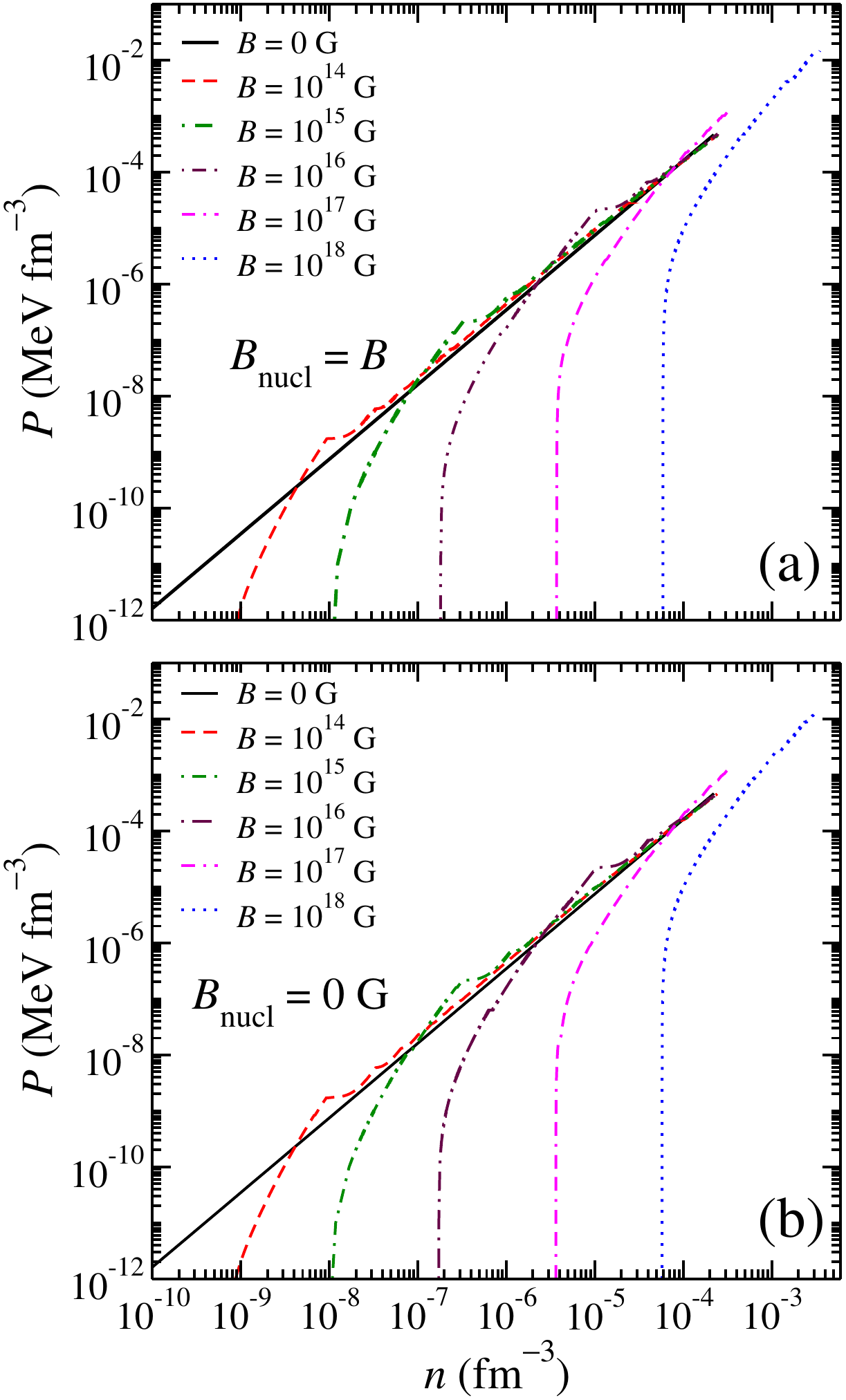}
\caption{(Color on-line) Equations of state ($n$ versus $P$) predicted by the DD-ME2 model, for six different external magnetic field values ($B=0 \, \mathrm{G}$, $B=10^{14} \, \mathrm{G}$, $B=10^{15} \, \mathrm{G}$, $B=10^{16} \, \mathrm{G}$, $B=10^{17} \, \mathrm{G}$ and  $B=10^{18} \, \mathrm{G}$), including ($B_{\rm nucl}=B$) and neglecting ($B_{\rm nucl}=0 \, \mathrm{G}$) the effects of magnetic fields on nuclei (respectively $a$ and $b$). Curves are plotted up to the neutron-drip transition point.}
 \label{fig:eos}
\end{figure}

In Fig.~\ref{fig:eos} is displayed the equation of state $P(n)$  as predicted by the DD-ME2 model, for six magnetic field values; for the NL3 model the behavior is similar. Panel (a) was obtained neglecting the effects of the magnetic field on nuclei, while on panel (b) they were fully taken into account. In the low density range, the magnetic field effect on the electron gas promotes a higher matter incompressibility, as signaled by the steep slope. As the pressure further increases $\nu \rightarrow \nu_{\mathrm{max}}$, and consequently the number of electronic levels which can be populated increases rapidly. Then, the magnetic field effects lose importance and the equation of state tends to approach the straight line associated to the $B=0$ case.

\section{Conclusions}
\label{conclusions}

For the first time the composition of the outer crust of a cold non-accreting and strongly magnetized neutron star has been studied taking into account, in a self-consistent fashion, the effects of the magnetic field on the binding energies of the nuclear species present in the Coulomb lattice. Results both including and neglecting the effect of the magnetic field on nuclear binding energies have been presented in order to understand its impact.

The required nuclear binding energies have been calculated using the NL3 and DD-ME2 relativistic mean-field models with explicit and fully self-consistent couplings to the magnetic field. These two models have been proven to be able to reproduce a wide variety of experimental data, including binding energies, with a reasonable accuracy.

Important changes in the structure and composition of the outer crust in the limit of high-magnetic fields ($B \sim 10^{16-18}$ G) have been found, even when neglecting its influence on nuclear binding energies. A shift to higher pressures, as the magnetic field increases, is observed in the transition from one nuclear species to another. The magnetic field effect on nuclear binding energies favors the appearance of heavier nuclei at low pressure, and, as the magnetic field increases, those heavier nuclei tend to be preferred up to greater pressures. In the most extreme case, and for the studied models, almost the whole outer crust is composed of ${}_{40}^{92}$Zr$_{52}$. 
Extreme magnetic fields also favor in some cases the appearance of odd-mass nuclei in the outer crust, in contrast to what happens for lower magnetic fields where the pairing interaction has a more relevant role. Furthermore, the neutron-drip transition pressure $P_{\mathrm{drip}}$ for the highest magnetic fields considered ($B > 10^{17} G$) is increased from one and up to two orders of magnitude with respect to the no magnetic field case. The latter will impact the spatial extension of the outer crust.

In summary, the inclusion of the effect of the magnetic field on nuclear binding energies may play a role in the determination of the outer crust composition, and thus in its mechanical and thermodynamical properties, for field strengths of about $10^{14}$ G (depending on the nuclear mass model used), and cannot be neglected for high field strengths ($B > 10^{16}$ G).

\begin{acknowledgments}
We are indebted to Isaac Vida\~na and Andrea Tiengo for useful discussions and a careful reading of the manuscript. 
\end{acknowledgments}

\appendix*

\section{numerical example}

\begin{table*}[h!]
\centering
\caption{Composition of the outer crust of a magnetar with $B=B_{\mathrm{nucl}}=10^{16} \, \mathrm{G}$. $P_{\mathrm{min}}$ ($P_{\mathrm{max}}$) is the minimum (maximum) pressure, in units of $\mathrm{MeV \, fm^{-3}}$, at which the given nucleus is present. The density $n_{\mathrm{max}}$, expressed in units of $\mathrm{fm^{-3}}$, is the maximum baryonic density at which the given nucleus is present, according to Eq.~\eqref{ne} and to charge neutrality hypotesis. $P_{\mathrm{ion}}$ is the electronic ionization pressure, which represents the lower limit of the outer crust.\\}
\begin{tabular*}{0.9\linewidth}{ccccccccccccc}

\toprule
\hline \\[-1.3ex]
Nucleus & $P_{\mathrm{min}}$ & $P_{\mathrm{max}}$ & $n_{\mathrm{max}}$ & $y$ & &Nucleus & $P_{\mathrm{min}}$ & $P_{\mathrm{max}}$ & $n_{\mathrm{max}}$ & $y$ \\ [0.5ex]
        & $\mathrm{MeV \, fm^{-3}}$&  $\mathrm{MeV \, fm^{-3}}$ & $\mathrm{fm^{-3}}$ & & &         & $\mathrm{MeV \, fm^{-3}}$&  $\mathrm{MeV \, fm^{-3}}$ & $\mathrm{fm^{-3}}$ & \\ [0.5ex]
\hline \\
\midrule
  \\[1.ex]
$\ce{^{62}_{28}Ni} $&$ P_{\mathrm{ion}} $&$ 1.43\cdot10^{-7} $&$ 9.02\cdot10^{-7} $&$ 0.432$   & ~~ &  $\ce{^{56}_{26}Fe} $&$ P_{\mathrm{ion}} $&$ 1.71\cdot10^{-7} $&$ 9.41\cdot10^{-7} $&$ 0.464$   \\[1.ex]
$\ce{^{88}_{38}Sr} $&$ 1.44\cdot10^{-7} $&$ 2.41\cdot10^{-6} $&$ 3.36\cdot10^{-6} $&$ 0.448$ &~~  & $\ce{^{88}_{38}Sr} $&$ 1.72\cdot10^{-7} $&$ 2.00\cdot10^{-6} $&$ 3.05\cdot10^{-6} $&$ 0.432$ \\[1.ex]
$\ce{^{87}_{37}Rb} $&$ 2.42\cdot10^{-6} $&$ 2.45\cdot10^{-6} $&$ 3.44\cdot10^{-6} $&$ 0.431$ &~~  & $\ce{^{87}_{37}Rb} $&$ 2.01\cdot10^{-6} $&$ 2.18\cdot10^{-6} $&$ 3.22\cdot10^{-6} $&$ 0.425$ \\[1.ex]
$\ce{^{68}_{28}Ni} $&$ 2.46\cdot10^{-6} $&$ 9.21\cdot10^{-6} $&$ 6.64\cdot10^{-6} $&$ 0.412$ &~~  & $\ce{^{86}_{36}Kr} $&$ 2.19\cdot10^{-6} $&$ 3.47\cdot10^{-6} $&$ 4.09\cdot10^{-6} $&$ 0.419$ \\[1.ex]
$\ce{^{84}_{34}Se} $&$ 9.22\cdot10^{-6} $&$ 2.02\cdot10^{-5} $&$ 9.96\cdot10^{-6} $&$ 0.409$ &~~  & $\ce{^{85}_{35}Br} $&$ 3.48\cdot10^{-6} $&$ 4.10\cdot10^{-6} $&$ 4.50\cdot10^{-6} $&$ 0.412$ \\[1.ex]
$\ce{^{77}_{30}Zn} $&$ 2.03\cdot10^{-5} $&$ 2.12\cdot10^{-5} $&$ 1.06\cdot10^{-5} $&$ 0.390$ &~~  & $\ce{^{84}_{34}Se} $&$ 4.11\cdot10^{-6} $&$ 1.73\cdot10^{-5} $&$ 9.23\cdot10^{-6} $&$ 0.405$ \\[1.ex]
$\ce{^{78}_{30}Zn} $&$ 2.13\cdot10^{-5} $&$ 2.94\cdot10^{-5} $&$ 2.23\cdot10^{-5} $&$ 0.385$ &~~  & $\ce{^{82}_{32}Ge} $&$ 1.74\cdot10^{-5} $&$ 2.36\cdot10^{-5} $&$ 1.60\cdot10^{-5} $&$ 0.390$  \\[1.ex]
$\ce{^{79}_{30}Zn} $&$ 2.95\cdot10^{-5} $&$ 4.09\cdot10^{-5} $&$ 2.89\cdot10^{-5} $&$ 0.380$ &~~  & $\ce{^{81}_{31}Ga} $&$ 2.37\cdot10^{-5} $&$ 2.73\cdot10^{-5} $&$ 2.07\cdot10^{-5} $&$ 0.383$ \\[1.ex]
$\ce{^{77}_{28}Ni} $&$ 4.10\cdot10^{-5} $&$ 8.62\cdot10^{-5} $&$ 5.65\cdot10^{-5} $&$ 0.363$&~~  & $\ce{^{80}_{30}Zn} $&$ 2.74\cdot10^{-5} $&$ 3.70\cdot10^{-5} $&$ 2.73\cdot10^{-5} $&$ 0.375$\\[1.ex]
$\ce{^{79}_{28}Ni} $&$ 8.63\cdot10^{-5} $&$ 1.44\cdot10^{-4} $&$ 8.16\cdot10^{-5} $&$ 0.354$&~~  & $\ce{^{78}_{28}Ni} $&$ 3.71\cdot10^{-5} $&$ 1.44\cdot10^{-4} $&$ 8.06\cdot10^{-5} $&$ 0.359$\\[1.ex]
$\ce{^{125}_{43}Tc} $&$ 1.45\cdot10^{-4} $&$ 1.57\cdot10^{-4} $&$ 9.61\cdot10^{-5} $&$ 0.344$&~~  &$\ce{^{122}_{40}Zr} $&$ 1.45\cdot10^{-4} $&$ 1.96\cdot10^{-4} $&$ 1.12\cdot10^{-4} $&$ 0.328$\\[1.ex]
$\ce{^{124}_{42}Mo} $&$ 1.58\cdot10^{-4} $&$ 1.74\cdot10^{-4} $&$ 1.05\cdot10^{-4} $&$ 0.339$ &~~  & $\ce{^{121}_{39}Y} $&$ 1.97\cdot10^{-4} $&$ 2.26\cdot10^{-4} $&$ 1.30\cdot10^{-4} $&$ 0.322$\\[1.ex]
$\ce{^{123}_{41}Nb} $&$ 1.75\cdot10^{-4} $&$ 1.90\cdot10^{-4} $&$ 1.13\cdot10^{-4} $&$ 0.333$&~~  & $\ce{^{120}_{38}Sr} $&$ 2.27\cdot10^{-4} $&$ 3.20\cdot10^{-4} $&$ 1.78\cdot10^{-4} $&$ 0.317$ \\[1.ex]
$\ce{^{122}_{40}Zr} $&$ 1.91\cdot10^{-4} $&$ 2.70\cdot10^{-4} $&$ 1.52\cdot10^{-4} $&$ 0.328$&~~  & $\ce{^{119}_{37}Rb} $&$ 3.21\cdot10^{-4} $&$ 3.35\cdot10^{-4} $&$ 1.86\cdot10^{-4} $&$ 0.311$\\[1.ex]
$\ce{^{121}_{39}Y} $&$ 2.71\cdot10^{-4} $&$ 3.08\cdot10^{-4} $&$ 1.70\cdot10^{-4} $&$ 0.322$&~~  & $\ce{^{117}_{35}Br} $&$ 3.36\cdot10^{-4} $&$ 4.30\cdot10^{-4} $&$ 2.37\cdot10^{-4} $&$ 0.299$\\[1.ex]
$\ce{^{120}_{38}Sr} $&$ 3.09\cdot10^{-4} $&$ 4.46\cdot10^{-4} $&$ 2.23\cdot10^{-4} $&$ 0.317$&~~ & $\ce{^{98}_{28}Ni} $&$ 4.31\cdot10^{-4} $&$ 5.19\cdot10^{-4} $&$ 2.78\cdot10^{-4} $&$ 0.286$\\[1.ex]
$\ce{^{119}_{37}Rb} $&$ 4.46.\cdot10^{-4} $&$ 4.78\cdot10^{-4} $&$ 2.44\cdot10^{-4} $&$ 0.311$&~~  & 
& & & & \\
\hline
\bottomrule
\label{tab:comp}
\end{tabular*}
\end{table*}

A complete sequence of the outer crust composition, taken into account the effect of the magnetic field on nuclei, is shown as an example in Table~\ref{tab:comp}. $P_{\mathrm{min}}$ ($P_{\mathrm{max}}$) is the minimum (maximum) pressure, in units of $\mathrm{MeV \, fm^{-3}}$, at which the given nucleus is present. The density $n_{\mathrm{max}}$, expressed in units of $\mathrm{fm^{-3}}$, is the maximum baryonic density at which the given nucleus is present, according to Eq.~\eqref{ne} and to charge neutrality. $P_{\mathrm{ion}}$ is the electronic ionization pressure, which represents the lower pressure limit of the outer crust  and $y$ is the proton fraction. The compositions correspond to the predictions of the DD-ME2 (left half side) and NL3 (right half side) models for a $B=10^{16}$ G case. It corresponds to the results in Fig.\ref{fig:ddme2}b and Fig.\ref{fig:nl3}b, respectively, depicted by using solid (blue) lines. 

%

\end{document}